\documentclass[reprint,amssymb, amsmath, aps, superscriptaddress, showpacs, footinbib,  prb]{revtex4-1}
\usepackage{graphicx}
\usepackage{color}

\newcommand{\eff}{\text{eff}}

\newcommand{\mg}{\text{mag}}

\begin{document}

\title{Hidden magnetic order in CuNCN}
\author{Alexander A. Tsirlin}
\email{altsirlin@gmail.com}
\affiliation{Max Planck Institute for Chemical Physics of Solids, N\"{o}thnitzer
Str. 40, 01187 Dresden, Germany}

\author{Alexander Maisuradze}
\affiliation{Laboratory for Muon Spin Spectroscopy, Paul Scherrer Institute, 5232 Villigen PSI, Switzerland}

\author{J\"org Sichelschmidt}
\author{Walter~Schnelle}
\author{Peter H\"ohn}
\affiliation{Max Planck Institute for Chemical Physics of Solids, N\"{o}thnitzer
Str. 40, 01187 Dresden, Germany}

\author{Ronald Zinke}
\author{Johannes Richter}
\affiliation{Institute for Theoretical Physics, University of Magdeburg, P.O. Box 4120,
39016 Magdeburg, Germany}

\author{Helge Rosner}
\email{Helge.Rosner@cpfs.mpg.de}
\affiliation{Max Planck Institute for Chemical Physics of Solids, N\"{o}thnitzer
Str. 40, 01187 Dresden, Germany}

\begin{abstract}
We report a comprehensive experimental and theoretical study of the quasi-one-dimensional quantum magnet CuNCN. Based on magnetization measurements above room temperature as well as muon spin rotation and electron spin resonance measurements, we unequivocally establish the localized Cu$^{+2}$-based magnetism and the magnetic transition around 70~K, both controversially discussed in the previous literature. Thermodynamic data conform to the uniform-spin-chain model with a nearest-neighbor intrachain coupling of about 2300~K, in remarkable agreement with the microscopic magnetic model based on density functional theory band-structure calculations. Using exact diagonalization and the coupled-cluster method, we derive a collinear antiferromagnetic order with a strongly reduced ordered moment of about 0.4~$\mu_B$, indicating strong quantum fluctuations inherent to this quasi-one-dimensional spin system. We re-analyze the available neutron-scattering data, and conclude that they are not sufficient to resolve or disprove the magnetic order in CuNCN. By contrast, spectroscopic techniques indeed show signatures of long-range magnetic order below 70~K, yet with a rather broad distribution of internal field probed by implanted muons. We contemplate the possible structural origin of this effect and emphasize peculiar features of the microstructure studied with synchrotron powder x-ray diffraction.
\end{abstract}

\maketitle

\section{Introduction}
\label{sec:intro}
While the majority of magnetic systems develops long-range order (LRO) of spins at low temperatures, the lack of LRO down to zero temperature is a less common scenario that implies strong quantum fluctuations and peculiar phenomena related to a specific type of spin lattice and/or exchange couplings. Low-dimensionality and frustration, two main prerequisites of quantum fluctuations, impede LRO in a quantum spin system by reducing the magnetic ordering temperature, the ordered magnetic moment, and other features of the ordered state.\cite{quantum-magnetism,frustrated} Although not necessarily preventing an ordered arrangement of spins at 0~K, quantum fluctuations render the LRO barely visible for experimental techniques that give response proportional to the size of magnetic moments (neutron scattering), transition entropy (specific heat), or other quantities related to the LRO state. For example, Sr$_2$CuO$_3$, a paradigmatic quasi-one-dimensional spin-$\frac12$ magnet, features the nearest-neighbor exchange coupling of $J\simeq 2600$~K,\cite{ami1995} yet undergoing LRO at as low as 5.4~K ($T_N/J\simeq 0.002$) with the ordered moment of 0.06~$\mu_B$, compared to 1~$\mu_B$ in a classical spin-$\frac12$ system.\cite{kojima1997} The LRO in Sr$_2$CuO$_3$ can only be observed with single-crystal neutron scattering\cite{kojima1997} or spectroscopic methods, such as muon spin rotation ($\mu$SR)\cite{keren1993,kojima1997} and nuclear magnetic resonance (NMR),\cite{thurber2001} while powder neutron diffraction essentially fails to detect the LRO because of the vanishingly small ordered moment.\cite{ami1995} 

The presence of a low-temperature LRO in some other low-dimensional spin systems remains a matter of contention. It is often declared that the lack of any visible anomalies in thermodynamic properties\cite{lancaster2007} as well as the lack of the magnetic neutron scattering in a powder experiment\cite{tsirlin2008} are a good evidence for the absence of LRO. However, $\mu$SR does find the magnetic transitions in many low-dimensional systems previously considered magnetically short-range-ordered in the experimentally accessible temperature range.\cite{lancaster2006,lancaster2007,pbvo3} A reliable study of the LRO in a low-dimensional spin system requires a comprehensive experimental investigation combined with a sound microscopic analysis that provides details of the LRO, such as the magnetic structure and the anticipated ordered moment. The microscopic input is essential for interpreting the experimental data, because the lack of the expected signal (e.g., in a powder neutron diffraction experiment) may either indicate the zero ordered moment (the absence of LRO), or simply show that the LRO state is beyond the sensitivity threshold of the method. In the following, we apply a combined experimental and theoretical approach to explore the ground state of CuNCN, a quasi-one-dimensional spin-$\frac12$ quantum magnet controversially discussed in the recent literature.

\begin{figure}
\includegraphics{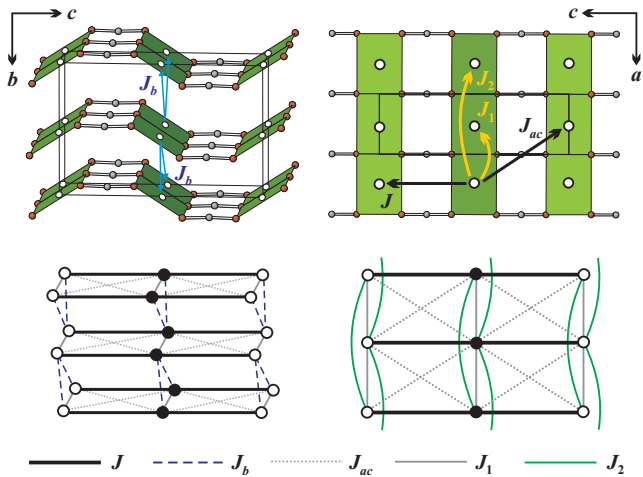}
\caption{\label{fig:structure}
(Color online) Top panels: crystal structure of CuNCN, open circles denote Cu atoms centering CuN$_4$ plaquettes (shaded). Bottom panels: microscopic magnetic model proposed in Ref.~\onlinecite{tsirlin2010}; open and filled circles at the sites of the spin lattice show up and down spins in the columnar AFM ground state (see Sec.~\ref{sec:theo}).
}
\end{figure}

CuNCN is a semiconducting compound containing spin-$\frac12$ Cu$^{+2}$ cations and anionic carbodiimide [NCN]$^{2-}$ groups. The layered crystal structure of CuNCN is formed by characteristic CuN$_4$ plaquettes that share edges and form chains along the $a$ direction. The nearly linear NCN groups connect the structural chains along $c$ (Fig.~\ref{fig:structure}).\cite{liu2005} An early experimental study by Liu \textit{et al.}\cite{liu2008} reported the lack of LRO, as concluded from thermodynamic measurements and neutron powder diffraction. The disordered ground state was understood as a ``highly correlated antiferromagnetic state at room temperature'' caused by the frustrated triangular spin lattice formed by the couplings $J_1$ and $J_b$.\cite{liu2008} Later, the same group performed a polarized neutron scattering experiment\cite{xiang2009} that did not show any signatures of LRO, either. However, this observation was now interpreted as a non-magnetic state of Cu$^{+2}$. In a subsequent theoretical study, Tchougr\'eeff and Dronskowski\cite{tchougreeff,*tchougreeff2} recalled the scenario of the magnetic Cu$^{+2}$ cations, and developed a mean-field theory for the resonating-valence-bond (RVB) ground state of an anisotropic $J_1-J_b$ triangular spin lattice (see the bottom left panel of Fig.~\ref{fig:structure}).

In a preceding study,\cite{tsirlin2010} we proposed an independent and essentially different microscopic picture,\cite{tsirlin2010} based on conventional density functional theory (DFT) band structure calculations that are well known as a highly efficient tool for unraveling complex spin lattices in quantum magnets.\cite{[{For example: }][{}]korotin1999,*rosner2002,*valenti2003,*dioptase} The leading antiferromagnetic (AFM) exchange $J\simeq 2500$~K was found along the $c$ direction via the NCN groups (Fig.~\ref{fig:structure}). A weaker FM coupling $J_1\simeq -500$~K connects the resulting spin chains along $a$, whereas a couple of marginal AFM couplings induce a weak frustration that should not preclude the LRO in CuNCN. In this paper, we provide experimental and further theoretical support for this microscopic scenario. Our data clearly indicate the presence of localized magnetic moments on the Cu$^{+2}$ atoms, and suggest the formation of static magnetic fields below 70~K. To understand the latter observation, we investigate the ground state of the proposed magnetic model, evaluate the N\'eel temperature, and explain why the anticipated magnetic order has likely been overlooked in the previous studies. Note that recently Zorko \textit{et al.}\cite{zorko} presented another experimental work on CuNCN and proposed an ``inhomogeneous magnetic ground state'' possibly related to the strong frustration on the triangular spin lattice. Although most of their experimental results match our findings, the interpretation is notably different. We discuss these differences in Sec.~\ref{sec:discussion} of the present paper and in a separate Comment.\cite{comment}

The outline of our paper is as follows. We start with methodological aspects in Sec.~\ref{sec:methods}, and proceed to the experimental results in Sec.~\ref{sec:experiment}. In Sec.~\ref{sec:ground}, we present a theoretical study of the microscopic magnetic model proposed in Ref.~\onlinecite{tsirlin2010}, and re-analyze the neutron-scattering data from Refs.~\onlinecite{liu2008,xiang2009}. We complete our work with a discussion and summary in Sec.~\ref{sec:discussion}.

\section{Methods}
\label{sec:methods}
Polycrystalline samples of CuNCN were prepared by a two-step procedure described in Ref.~\onlinecite{liu2005}. We used CuCl$_2\cdot 2$H$_2$O and H$_2$NCN as starting materials and Na$_2$SO$_3$ as a reducing agent. H$_2$NCN was stored in a fridge to avoid polymerization. CuNCN was obtained in the form of black powder. Repeated attempts of crystal growth were so far unsuccessful. Although laboratory powder x-ray diffraction (XRD, see below) showed similar patterns for all samples, further studies (chemical analysis, synchrotron XRD, magnetization measurements) revealed deviations in the chemical composition, microstructure, and magnetic behavior.

Powder samples of CuNCN were characterized with laboratory XRD (Huber G670 Guinier camera, CuK$_{\alpha1}$ radiation, Ge monochromator, ImagePlate detector, $2\theta=3-100^{\circ}$ angle range) and conventional analytical techniques.\footnote{%
The Cu content was analyzed by inductively-coupled-plasma atomic-emission spectroscopy (ICP-AES). Nitrogen and oxygen as well as carbon and hydrogen content was determined from carrier-gas hot extraction/combustion methods, respectively.}
High-resolution XRD patterns for selected samples were measured at room temperature at the ID31 beamline of European Synchrotron Radiation Facility in Grenoble (wavelength $\lambda=0.4$~\r A, eight scintillation detectors preceded by Si (111) analyzer crystals, $2\theta=1-40^{\circ}$ angle range). Samples were loaded into thin-walled borosilicate glass capillaries and spun during the data collection.

Magnetic susceptibility was measured with a Quantum Design MPMS SQUID magnetometer in the temperature range $2-650$~K in applied fields up to 7~T. Heat capacity was studied by a relaxation technique (Quantum Design PPMS) in the temperature range $2-100$~K in zero magnetic field.

The electron spin resonance (ESR) measurements were performed with a standard continuous-wave spectrometer at X-band frequencies ($\nu\approx 9.5$~GHz) by using a cylindrical resonator in TE$_{012}$ mode. The ESR spectra were analyzed by fitting them with Lorentzian lines. From these fits, the linewidth $\Delta B$ (half-width at half-maximum), the resonance field $B_{res}$ (providing the ESR $g$-factor), and the intensity (given by the area $\propto A\Delta B^{2}$ under the ESR absorption, where $A$ is the amplitude of the Lorentzian) were determined.

Zero- and longitudinal-field (ZF\&LF) $\mu$SR experiments were performed at the $\pi$M3 beam line at the Paul Scherrer Institute on the GPS spectrometer (Villigen, Switzerland). Forward, backward, up, and down positron detectors were used for monitoring 
the asymmetry signal $A(t)$. The ZF measurements were performed in the transverse mode with muon spin perpendicular to 
its momentum, while the LF measurements were done in the LF mode with the muon spin parallel to its momentum. The ZF spectra were obtained in the temperature range of $5-200$~K, while the LF measurements were done in a series of fields ranging from 1~mT to 640~mT at 100 K and at 30 mT in the temperature range of 5 to 250 K. Typical counting statistics were $\sim5\times10^6$ positron events per each particular data point.

The ground state of the proposed microscopic magnetic model was studied by coupled-cluster and Lanczos diagonalization calculations using the program packages ``The crystallographic CCM'' (by D.J.J. Farnell and J. Schulenburg) and J. Schulenburg's {\it spinpack}, respectively. To evaluate the N\'eel temperature, we simplified our spin model (see Sec.~\ref{sec:theo}) and  performed quantum Monte Carlo simulations using the \texttt{loop} algorithm\cite{looper} of the ALPS package.\cite{alps} Neutron-scattering patterns for possible magnetic structures of CuNCN were calculated with \texttt{FullProf}.\cite{fullprof}

\section{Experimental results}
\label{sec:experiment}

\subsection{Magnetic properties}
\label{sec:magnetic}
Liu \textit{et al.}\cite{liu2008} reported magnetic susceptibility ($\chi$) measurements for CuNCN up to 320~K. They observed a nearly temperature-independent susceptibility above 100~K, whereas at lower temperatures the data showed a bend around 70~K followed by a Curie-like paramagnetic upturn. Additionally, the increase in the magnetic field induced a systematic reduction in $\chi$, thereby indicating a minor ferromagnetic contribution of unknown origin. We extended the measurements by Liu \textit{et al.}\cite{liu2008} in two aspects. First, we collected the data above 320~K, since there is strong evidence for large couplings on the order of 2000~K,\cite{tsirlin2010} so that an evaluation of exchange parameters would require the high-temperature data. Second, we performed the measurements for several samples to sort out external contributions and to better understand the role of impurities. We start with the high-temperature data, while the sample dependence will be discussed in Sec.~\ref{sec:samples}.

\begin{figure}
\includegraphics[width=6.5cm,angle=90]{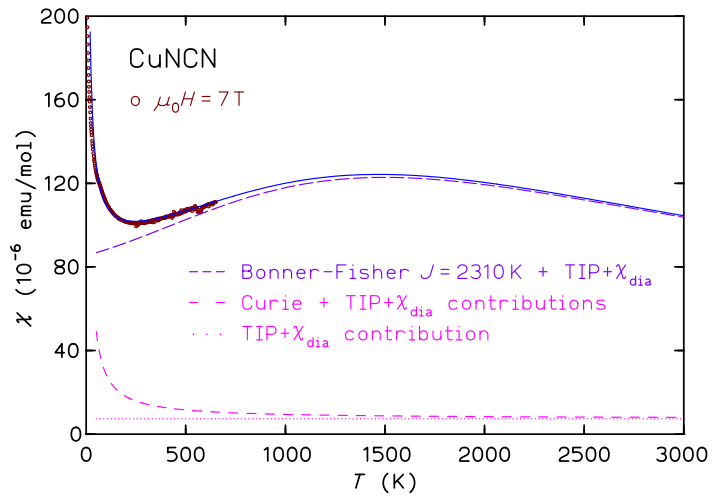}
\caption{\label{fig:fit}
(Color online) Magnetic susceptibility of CuNCN measured up to 650~K in the applied field of 7~T (open circles) and the fit with Eq.~\eqref{eq:fit} (solid line) along with individual contributions described in the text (dashed and dotted lines).
}
\end{figure}

Fig.~\ref{fig:fit} presents the susceptibility data collected up to 650~K, which is close to the decomposition temperature of CuNCN.\cite{foot1} To check for the stability of the sample in this temperature range, we performed the magnetization measurements both on heating and on cooling. Both datasets perfectly matched, thus confirming the intrinsic nature of the observed signal. Above 250~K, the magnetic susceptibility of CuNCN increases. This contradicts the itinerant scenario requiring an essentially temperature-independent Pauli paramagnetism, and challenges the conclusion of Ref.~\onlinecite{xiang2009} on the non-magnetic nature of Cu$^{+2}$ cations in CuNCN. The increase in $\chi$ is a clear signature of localized spins that are subject to strong AFM couplings inducing a low-dimensional and/or frustrated behavior. Quantum fluctuations in low-dimensional/frustrated magnets typically lead to a broad susceptibility maximum. In our case, the maximum lies above the accessible temperature range\cite{foot1} (Fig.~\ref{fig:fit}), and the measurement can only probe the susceptibility below the maximum. Our measurements are, therefore, high-temperature on the usual experimental scale, but cover a relatively narrow temperature range on the scale of the leading exchange coupling $J\simeq 2500$~K ($T/J\leq 0.3$).

Although the data in the narrow temperature range can hardly be used to establish details of the magnetic model, it is still possible to fit the experimental susceptibility using a given model, and evaluate the relevant exchange couplings. Since DFT calculations\cite{tsirlin2010} consistently find the leading AFM coupling $J\simeq 2500$~K along the $c$ direction, we use the model of a uniform Heisenberg spin-$\frac12$ chain supplemented by a temperature-independent contribution $\chi_0$ for core diamagnetism/van Vleck paramagnetism as well as a Curie contribution $C/T$ for the low-temperature upturn related to the non-interacting impurity spins:
\begin{equation}
  \chi=\chi_0+\dfrac{C}{T}+\dfrac{N_Ag^2\mu_B^2}{J}\chi_{\text{chain}}(T/J).
\label{eq:fit}\end{equation}
Here, $g$ is the $g$-factor, $N_A$ is Avogadro's number, $\mu_B$ is Bohr magneton, and $\chi_{\text{chain}}(T/J)$ is the susceptibility of a uniform Heisenberg chain, as given in Ref.~\onlinecite{johnston2000}. Our fit yields $\chi_0=7.3\times 10^{-6}$~emu/mol, $g=2.2$, $C=0.0021$~emu~K/mol, and $J=2310$~K. The fitted $g$-value is in good agreement with the ESR estimate $g\simeq 2.1$ (see Sec.~\ref{sec:esr}),\footnote{%
The marginal difference in the $g$-factor likely originates from the predominantly FM interchain couplings ($J_1\simeq -500$~K) that are not considered in the fitting expression of Eq.~\eqref{eq:fit}. Such couplings would increase the susceptibility maximum and therefore decrease the fitted $g$ value.}
whereas our experimental estimate of $J$ almost perfectly matches the computational prediction $J\simeq 2500$~K of Ref.~\onlinecite{tsirlin2010}. This result provides a strong experimental support for the proposed microscopic magnetic model.

\begin{figure}
\includegraphics{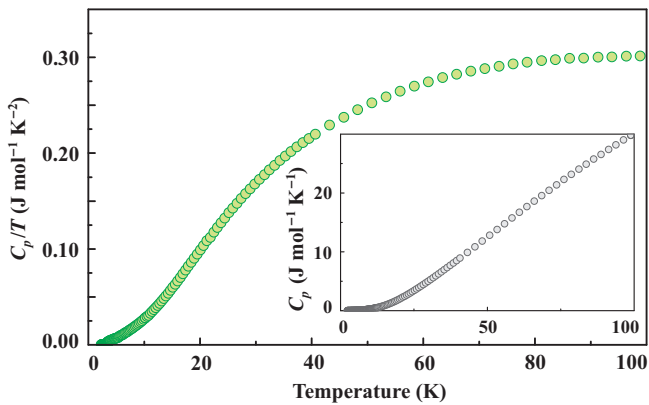}
\caption{\label{fig:heat}
(Color online) Specific heat of CuNCN measured in zero magnetic field.
}
\end{figure}

An important consequence of our model is the low-temperature LRO driven by interchain couplings (see also Sec.~\ref{sec:ground}).\cite{tsirlin2010} The bend of the susceptibility curve around 70~K (Fig.~\ref{fig:chi}, upper panel) is a possible signature of such ordering. However, the intrinsic nature of the susceptibility anomaly has been questioned by Liu \textit{et~al.},\cite{liu2008} because any visible effect in the specific heat ($C_p$) is lacking. Fig.~\ref{fig:heat} shows that the heat capacity of CuNCN is indeed smooth up to 100~K, but this observation does not evidence against the LRO. In Sec.~\ref{sec:intro}, we mentioned that the ordering transitions in quantum magnets may be invisible for thermodynamic measurements because of the diminutively small entropy release at the low transition temperatures $T_N/J\simeq 0.03$ (see also Sec.~\ref{sec:theo}). This is the case for CuNCN, where local probes give clear indications for the formation of static magnetic fields below $70-80$~K (see Sections~\ref{sec:musr} and~\ref{sec:esr}).

\subsection{Sample characterization}
\label{sec:samples}
\begin{figure}
\includegraphics{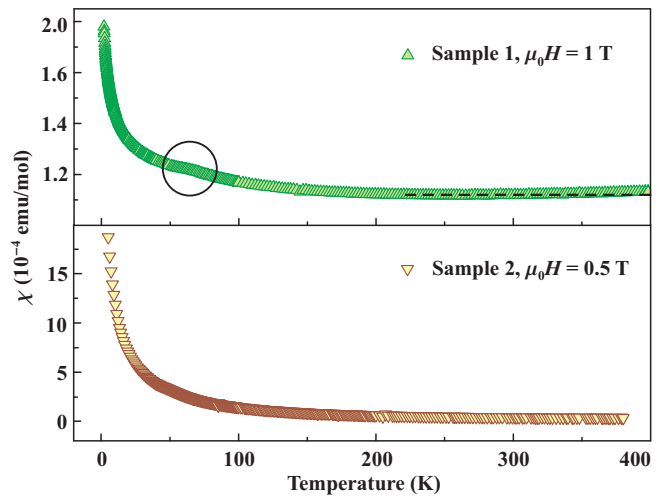}
\caption{\label{fig:chi}
(Color online) Magnetic susceptibility for two samples of CuNCN with different chemical composition (Table~\ref{tab:composition}), different microstructure (Fig.~\ref{fig:xrd}), and different concentration of paramagnetic impurities (see text for details). The dashed line in the upper panel denotes the minimum susceptibility and underscores the weak increase in $\chi$ above 300~K (compare to the high-temperature data in Fig.~\ref{fig:fit}). The circle marks the bend of the curve around 70~K.
}
\end{figure}

Prior to discussing the $\mu$SR and ESR results, we will comment on the problem of possible impurities in the polycrystalline samples of CuNCN. The comparative study of several samples demonstrated that the magnitude of the low-temperature Curie-like impurity contribution is strongly sample-dependent. For a further study, we selected samples 1 and 2 showing the least pronounced and most pronounced low-temperature upturns, respectively (Fig.~\ref{fig:chi}). Fitting the low-temperature susceptibility with the Curie law, we roughly estimate the amount of an effective spin-$\frac12$ impurity as 0.1~\% for sample 1 and 1.3~\% for sample 2. It is worth noting that the bend around 70~K, which is sometimes thought to be of impurity origin,\cite{liu2008,xiang2009} is in fact more pronounced in sample 1 with the significantly lower impurity contribution. In sample 2, the anomaly is masked by the stronger impurity signal.

The powder XRD suggests that both samples 1 and 2 are single-phase, although this method is sensitive to crystalline phases, only. To test the bulk composition of the samples, we used the chemical analysis. Representative results are reported in Table~\ref{tab:composition}. Sample 1 is close to the nominal composition, whereas sample 2 considerably deviates from the anticipated CuNCN composition and contains appreciable amounts of oxygen and hydrogen. While the deviations in the composition of sample 2 could be ascribed to an amorphous impurity phase invisible for XRD, the actual situation is more complex. The two samples not only differ in the chemical composition, but also contain CuNCN-type phases with different features of the microstructure. The differences in the crystalline components of samples 1 and 2 are evidenced by high-resolution XRD, and further confirmed by the $\mu$SR and ESR studies (Sections~\ref{sec:musr} and~\ref{sec:esr}, respectively).

\begin{figure}
\includegraphics{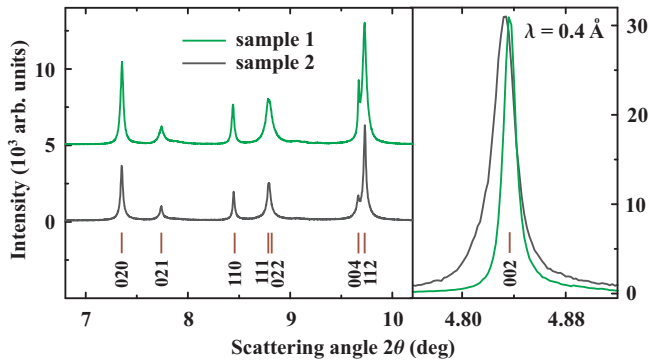}
\caption{\label{fig:xrd}
(Color online) Synchrotron XRD patterns for two samples of CuNCN with different chemical composition. In the left panel, the pattern of sample 1 is offset for +5000 for clarity. Sample 1 shows a pronounced anisotropy of reflection halfwidth (note the 004 and 112 reflections in the left panel), whereas the reflections of sample 2 demonstrate more isotropic angular dependence of the halfwidth and peculiar reflection shape (right panel).
}
\end{figure}

\begin{table}
\caption{\label{tab:composition}
Chemical composition (in~wt.\%) of the CuNCN samples.
}
\begin{ruledtabular}
\begin{tabular}{cccccc}
            & Cu   & N     & C    & H   & O   \\
   Sample 1 & 60.8 & 27.1  & 11.6 & 0.1 & 0.0 \\
   Sample 2 & 59.4 & 24.5  & 11.1 & 0.4 & 4.7 \\
   Nominal  & 61.4 & 27.0  & 11.6 & 0.0 & 0.0 \\
\end{tabular}
\end{ruledtabular}
\end{table}
To better characterize the crystalline CuNCN phase in samples 1 and 2, we performed XRD measurements using a synchrotron source that provides an excellent resolution and shows remarkable sensitivity to the microstructure. The room-temperature powder patterns of samples 1 and 2 are shown in Fig.~\ref{fig:xrd}. In sample 1, we observed a dramatic anisotropy of the reflection halfwidth. The reflections $hkl$ with both $k$ and $l$ non-zero are systematically broadened, as seen, e.g., from a comparison of the two neighboring reflections, 004 and 112 at $2\theta\simeq 9.7^{\circ}$, or 021 and 110 at lower $2\theta$. Reflections of sample 2 look more isotropic in terms of the halfwidth, yet their shape is somewhat unusual (see the left panel of Fig.~\ref{fig:xrd}) and precludes the conventional Rietveld refinement. We find that both samples 1 and 2 are not perfectly crystallized and exhibit peculiar features of the microstructure. 

At this point, we also comment on the sample characterization reported in previous studies of CuNCN.\cite{liu2005,liu2008,zorko} In sample 1, the chemical composition as well as the magnitude of the Curie-like susceptibility upturn are compatible with the data of Refs.~\onlinecite{liu2005,liu2008,zorko}, thereby indicating similar sample quality. Our sample 2 may be thought to be of lower quality, although the deviations in its chemical composition coexist with -- and probably signal -- a peculiar change in the microstructure of the crystalline CuNCN phase. We argue that the previously overlooked problem of defects and microstructure might be of crucial importance for understanding the magnetism. Although the authors of Refs.~\onlinecite{liu2005,zorko} describe their samples of CuNCN as having ``excellent crystallinity'' evidenced by noticable high-angle Bragg peaks in laboratory x-ray data obtained with CuK$_{\alpha1}$ radiation, this test is not sensitive to extended defects observed in our study. Similar to Ref.~\onlinecite{liu2005}, the synchrotron data for samples 1 and 2 show a number of sharp reflections up to $q\simeq 7.5$~\r A$^{-1}$ ($2\theta\simeq 133^{\circ}$ for CuK$_{\alpha1}$ radiation), but this neither precludes the dramatic broadening of the low-angle reflections with non-zero $k$ and $l$ nor excludes the peculiar reflection shape observed in sample~2. 

The variable crystallinity of CuNCN is unrelated to the particle size, which would lead to a uniform broadening of all Bragg peaks, but rather concerns specific defects resolvable in a high-resolution XRD experiment, only. For example, the powder XRD data reported in Refs.~\onlinecite{liu2005,zorko} show the 004 and 112 reflections at $2\theta\simeq 9.7^{\circ}$ as a single peak, and do not allow to observe the difference in the halfwidths of these reflections. Therefore, a further characterization with high-resolution XRD is desirable. We believe that genuine ``powder samples of CuNCN with excellent crystallinity'' remain a challenging topic, and the problem of microstructure requires further careful consideration that lies beyond the scope of the present study.

\subsection{$\mu$SR spectroscopy}
\label{sec:musr}

\begin{figure}
\includegraphics{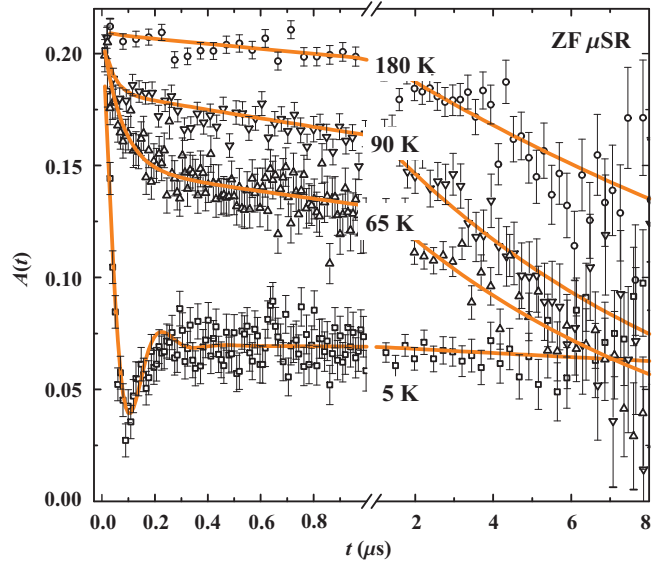}
\caption{\label{fig:muSR_t_spectra}
(Color online) ZF $\mu$SR time spectra of sample 1 measured at 5, 65, 90, and 180 K. The solid lines are fits 
to the data with Eq.~\eqref{eq:muSRfitModel}. 
}
\end{figure}

Figure~\ref{fig:muSR_t_spectra} shows ZF $\mu$SR time spectra of sample 1 measured at 5, 65, 90, and 180~K. At 5~K, a strongly depolarized $\mu$SR signal is evident, suggesting a magnetically ordered state of the sample. With increasing temperature, the muon depolarization gradually vanishes, and a low relaxation rate at 180~K identifies a complete transition to the paramagnetic state.
\begin{figure}
\includegraphics{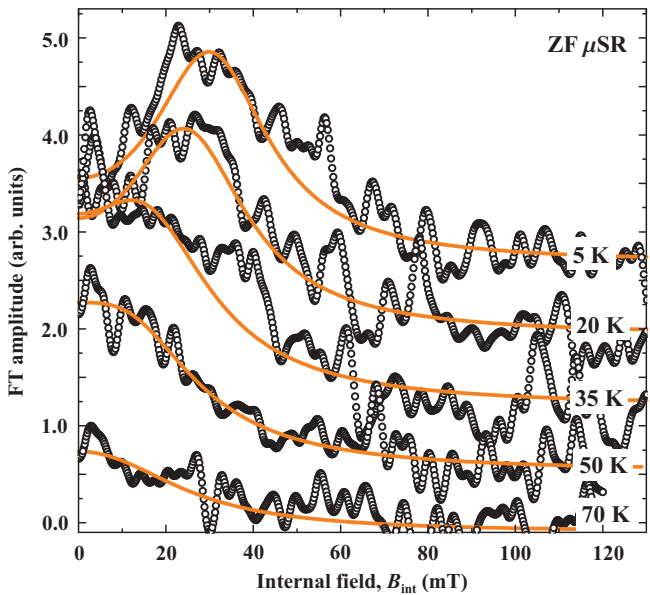}
\caption{\label{fig:muSR_FT}
(Color online) Fourier transform (FT) of the oscillating (2/3) fraction of the $\mu$SR time spectra measured at 5, 20, 35, 50, and 70~K. The solid lines are FT of the oscillating fraction from fits to the data with Eq.~\eqref{eq:muSRfitModel}. The non-zero first moment of the internal field distribution is evident at 5~K and gradually vanishes above 50~K. For better visualization, the FT spectra below 70~K are shifted vertically. }
\end{figure}
\begin{figure*}
\includegraphics{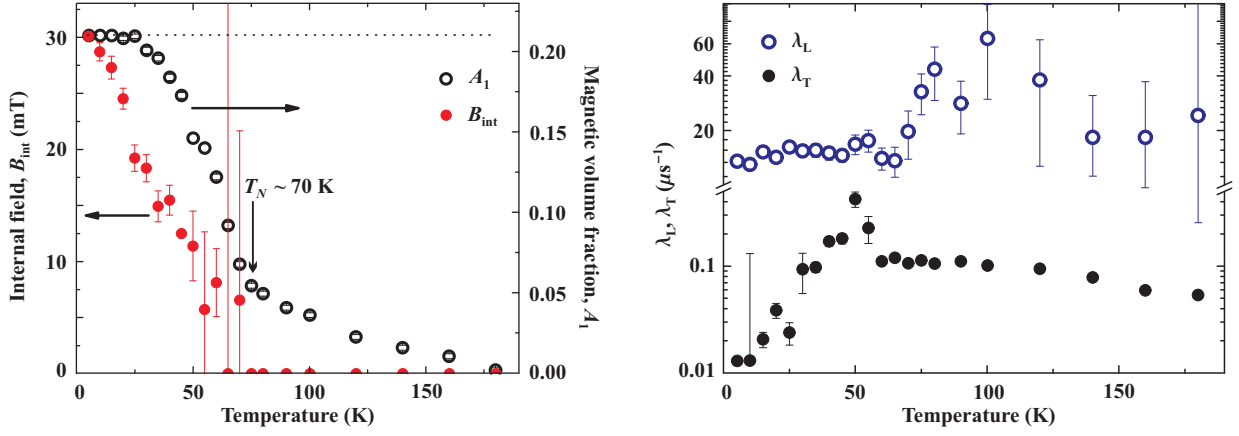}
\caption{\label{fig:muSR_ZF_results}
(Color online) Summary of ZF $\mu$SR results fitted with Eq.~\eqref{eq:muSRfitModel}. Left panel: temperature dependence of magnetic volume fraction $A_1$ and internal magnetic field $B_{\rm int}$. Right panel: longitudinal ($\lambda_{\rm L}$) and transverse ($\lambda_{\rm T}$) relaxation rates as a function of temperature.}
\end{figure*}%
The time dependence of the ZF $\mu$SR time spectra is well described with the following equation:\cite{Blundell99muSR}
\begin{align}\label{eq:muSRfitModel}
A(t) =& A_1\left[\frac{2}{3}\exp(-\lambda_{\rm T} t)\cos(\gamma_{\mu}B_{\rm int} t) +
\frac{1}{3}\exp(-\lambda_{\rm L} t)\right] \\
    & + A_2 \exp(-\lambda_2 t).
\end{align}
The first term gives the signal of the magnetically-ordered polycristalline sample, with 2/3 oscillating and 1/3 non-oscillating parts, while the second term describes the paramagnetic fraction of the sample. Initial asymmetries $A_1$ and $A_2$ are proportional to the volume fractions of the magnetically-ordered and paramagnetic phases, respectively, and add up to the temperature-independent total asymmetry $A_1+A_2=0.21$. The transverse relaxation rate $\lambda_{\rm T}$ is proportional to the dispersion of the internal magnetic field $B_{\rm int}$. $\lambda_{\rm L}$ is the longitudinal relaxation rate of muon polarization, whereas $\gamma_\mu=2\pi\cdot135.53$~MHz/T is the muon gyromagnetic ratio. The parameter $\lambda_2$ describes the muon relaxation rate in the paramagnetic fraction of the sample. It is related to the static and dynamic local fields.\cite{Pratt07} 

In contrast to the recent study by Zorko {\it et~al.},\cite{zorko} we clearly identify a nonzero internal field at 5~K in sample 1. Fig.~\ref{fig:muSR_FT} shows the Fourier transform (FT) for the oscillating fraction (2/3) of the $\mu$SR-time spectrum (i.e., the paramagnetic fraction and the non-oscillating (1/3) fraction of the signal were subtracted from the full $\mu$SR time-spectrum). The non-zero internal field is a direct proof for the antiferromagnetically ordered rather than a spin-glass state of stoichiometric CuNCN. We note, however, the broad distribution of internal fields that may be related to several inequivalent muon stop sites in the structure and/or to a structural inhomogeneity. This problem is further discussed in Sec.~\ref{sec:discussion}.

Figure~\ref{fig:muSR_ZF_results} shows the ZF $\mu$SR results for $B_{\rm int}$, $\lambda_{\rm T}$, $\lambda_{\rm L}$, and $A_1$  evaluated using Eq.~\eqref{eq:muSRfitModel}. The relaxation rate of the paramagnetic fraction $\lambda_2$ is zero below 60~K and very close to $\lambda_{\rm L}$ above 60~K. Therefore, we set $\lambda_2=\lambda_{\rm L}$ above 60~K to avoid correlations between these parameters. At 5~K, 100~\% volume fraction of the sample is in the AF ordered state (see $A_1(T)$ in the left panel of Fig.~\ref{fig:muSR_ZF_results}). With increasing the temperature above 25~K, the magnetic volume fraction gradually decreases to 50\% at about 62~K and to $\simeq 25$\% at 70~K. At higher temperatures, $A_1$ decreases linearly and eventually vanishes at 180~K. By contrast, the internal magnetic field $B_{\rm int}$ probed at the muon site(s) decreases with temperature from 30~mT at 5~K to 0 at $\simeq 70$~K. Above 70~K, spins are in a highly disordered state, as seen from the temperature dependence of the transverse relaxation rate $\lambda_{\rm T}$ (right panel of Fig.~\ref{fig:muSR_ZF_results}). The longitudinal relaxation rate also shows a small peak at about 50~K, as expected at the proximity to a magnetic transition. This peak is smeared due to a rather broad magnetic transition. Presently, we are unable to find a unique reason for the marginal fraction of the ordered phase (non-zero $A_1$) between 70 and 180~K. Plausible explanations would be the formation of a muon-induced magnetic phase in small regions of the sample or the structure inhomogeneity. The zero internal field as well as the well-defined ESR absorption above 70~K (Sec.~\ref{sec:esr}) corroborate the lack of the macroscopic LRO above 70~K.

\begin{figure}
\includegraphics{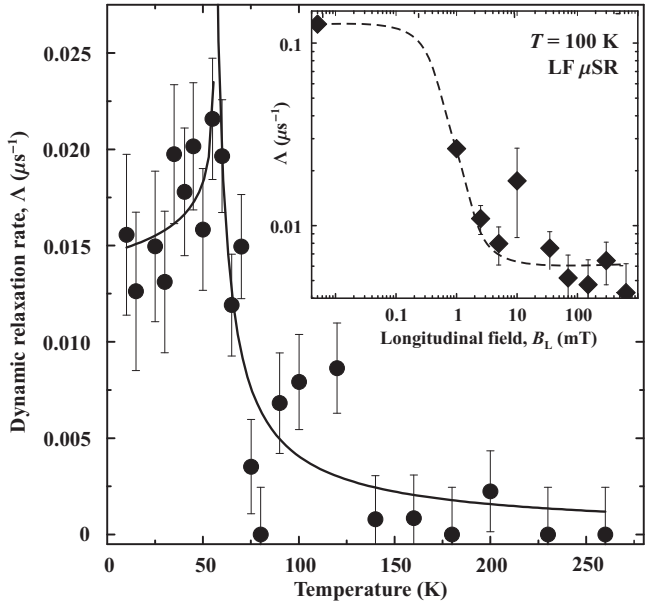}
\caption{\label{fig:muSR_LF_results}
Temperature dependence of the dynamic relaxation rate $\Lambda$ as obtained with Eq.~\eqref{eq:muSR_lf_model} in the field of $B_{\rm L}=30$ mT. The inset shows the decoupling LF experiment at 100 K. The solid and dashed lines are guides to the eye.}
\end{figure}

The $\mu$SR time-spectrum of the paramagnetic fraction is described with an exponential function [see Eq.~\eqref{eq:muSRfitModel}]. The exponential relaxation rate is characteristic either for the dynamic muon relaxation or for the relaxation due to static diluted magnetic centers. In order to distinguish between these two possibilities, we performed the LF $\mu$SR experiment (also referred as a decoupling experiment) at 100~K where the dominant fraction of the sample is in the paramagnetic state. The $\mu$SR time spectra were analyzed with the equation: 
\begin{equation}\label{eq:muSR_lf_model}
A(t) = A'\exp(-\Lambda t).
\end{equation} 
The fit was performed for the times above 0.5 $\mu$s to exclude the influence of the fast relaxing component of the $\mu$SR time-spectrum (see Fig.~\ref{fig:muSR_t_spectra}). The longitudinal-field ($B_{\rm L}$) dependence of $\Lambda$ is shown in the inset of Fig.~\ref{fig:muSR_LF_results}. The $\mu$SR signal is decoupled at about 0.5~mT suggesting the dominant static character of muon depolarization in the paramagnetic state. Therefore, to study the weak dynamic muon relaxation we performed a temperature scan in the applied longitudinal field $B_{\rm L} = 30$~mT where static internal fields are completely decoupled. The fitting results are summarized in Fig.~\ref{fig:muSR_LF_results}. The maximum relaxation rate $\Lambda$ is 0.02~$\mu$s$^{-1}$ which is close to the instrument resolution limit resulting in substantial error. The temperature dependence of $\Lambda$ has a peak at about 50~K, similar to one observed on $\lambda_{\rm L}$ in ZF $\mu$SR (right panel of Fig.~\ref{fig:muSR_ZF_results}).

\begin{figure}
\includegraphics{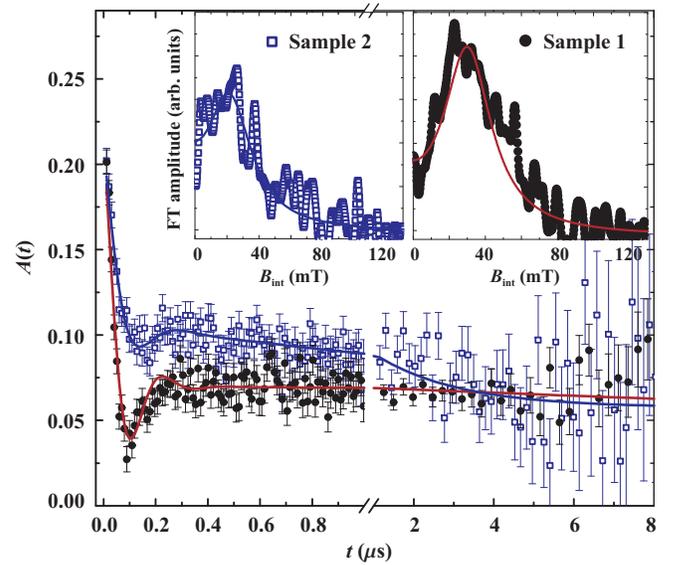}
\caption{\label{fig:muSR_samples}
(Color online) ZF $\mu$SR spectra of samples 1 and 2 measured at 5~K and the respective Fourier transforms showing internal fields $B_{\rm int}$.}
\end{figure}

We also measured the ZF $\mu$SR spectrum for the off-stoichiometric sample 2 at 5~K (Fig.~\ref{fig:muSR_samples}). In contrast to sample 1, sample 2 retains a weak paramagnetic component down to low temperatures, and shows a more shallow dip in $A(t)$. The latter feature indicates a broader distribution of internal fields, as confirmed by the Fourier transforms shown in the insets of Fig.~\ref{fig:muSR_samples}. In sample 1, the distribution of $B_{\rm int}$ has a well-resolved maximum around 30~mT, whereas in sample 2 internal fields between 0 and 20~mT are observed with nearly equal probabilities. The different $\mu$SR spectra confirm that the primary differences between the samples are unrelated to the possible contamination with an amorphous paramagnetic impurity, but rather reflect the differences in the crystalline ``CuNCN'' phase.
\subsection{ESR}
\label{sec:esr}
\begin{figure*}
\includegraphics[width=15cm]{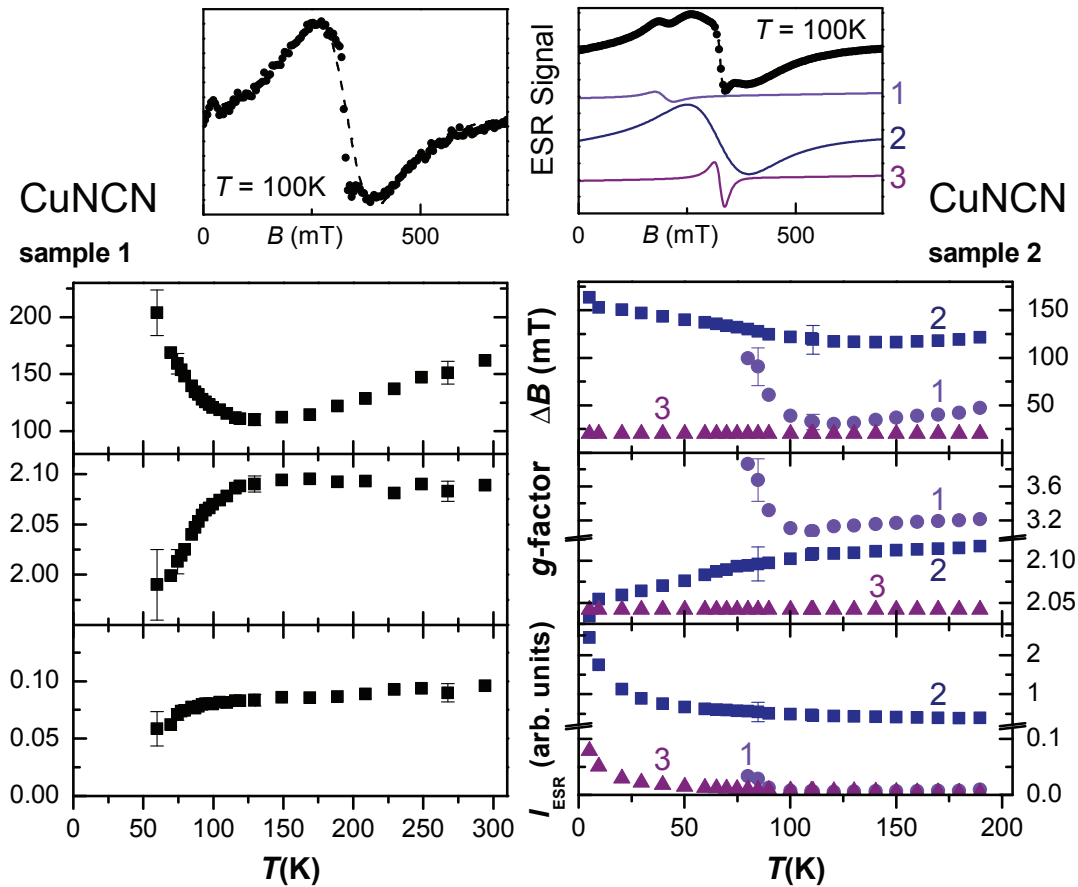}
\caption{\label{fig:esr}
(Color online) X-band ESR results of CuNCN (sample 1, left panels and sample 2, right panels). The ESR signals (field derivative of absorbed microwave power) were fitted with a Lorentzian shape (dashed lines) using a two lines (sample 1) and three lines (sample 2), as further explained in the text. The lower panels display the corresponding parameters linewidth ($\Delta B$), $g$-factor (determined by the resonance field) and intensity (integrated microwave absorption). 
}
\end{figure*}

While muon scattering directly probes the amount of the magnetically-ordered phase, this experiment involves the interaction between the crystalline material and implanted muons. Therefore, extrinsic phenomena driven by muons and, particularly, the local magnetic order induced by muons can not be excluded. To check for the intrinsic nature of the observed magnetic transition, we performed the ESR measurements. 

The ESR spectra of the two samples of CuNCN (Fig.~\ref{fig:esr}) show clear differences except for the narrow line with the temperature-independent linewidth of 20~mT at a field of 327~mT ($g=2.04(2)$). These parameters are typically expected for the resonance of non-interacting Cu$^{+2}$ spins. The left panel of Fig.~\ref{fig:esr} shows the ESR spectrum of sample 1 which consists, beside the narrow line, of one broad Lorentzian line denoted by the dashed curve in the experimental spectrum. The parameters of this broad line display a temperature dependence characteristic to the presence of the low-temperature magnetic order. The pronounced increase (decrease) in the linewidth ($g$-factor) indicates the slowing down of spin fluctuations and the onset of internal magnetic fields in agreement with the $\mu$SR observations. Upon cooling, the line intensity is slowly decreasing and eventually drops down to zero below 70~K. Therefore, the intrinsic magnetic susceptibility of CuNCN vanishes at low temperatures, as typical for the LRO state. Another possible interpretation is the presence of a spin gap that leads to the decrease in the magnetic susceptibility at low temperatures and, consequently, to the vanishing ESR line. Overall, the temperature evolution of the two lines in the ESR spectrum is consistent with the magnetic susceptibility of sample 1 (Fig.~\ref{fig:chi}, upper panel). Based on the observation of two different lines, we are able to separate the experimental susceptibility into the Curie-like paramagnetic contribution and the intrinsic contribution of CuNCN that vanishes below 70~K.
 
The ESR spectra of sample 2 could be fitted by three Lorentzians as shown by the solid lines in the upper right panel of Fig. \ref{fig:esr}. Line 3 is the aforementioned narrow temperature-independent line. This line is superimposed by a strong and broad line 2 dominating the spectra in the whole investigated temperature range. In agreement with the larger susceptibility of sample 2 (Fig.~\ref{fig:chi}), this line causes an order of magnitude larger ESR intensity for sample 2 compared to sample 1. Despite similar linewidths, line 2 in sample 2 is rather different from the broad line in sample 1. While the latter disappears below 70~K, the former is observed down to 5~K, with marginal changes in the linewidth and $g$-value. Interestingly, the ESR spectrum of sample 2 features an additional contribution (line 1) that does vanish below 70~K, yet showing a smaller linewidth and smaller resonance field (higher $g$-factor) than the intrinsic signal (broad line) in sample 1. Therefore, the temperature of 70~K is also characteristic for sample 2, although the magnetic transition (or the spin gap) takes place in a part of the sample only. This compares well to the incomplete and rather inhomogeneous magnetic ordering observed with $\mu$SR (Fig.~\ref{fig:muSR_samples} and Sec.~\ref{sec:musr}). 

Presently, we are unable to decide whether lines 1 and 2 indicate different magnetic phases of sample 2, or evidence some peculiar magnetism within the single off-stoichiometric CuNCN phase. However, both ESR and $\mu$SR prove that sample 2 can not be simply considered as the crystalline CuNCN phase mixed with an amorphous impurity. The deviations in the chemical composition and/or microstructure have strong influence on the magnetism of CuNCN.

\section{Theoretical analysis}
\label{sec:ground}
The ESR and $\mu$SR results give strong evidence for the emergence of static magnetic fields in CuNCN below 70~K. This onset temperature matches the weak susceptibility anomaly. However, the specific heat and neutron-scattering data\cite{liu2008,xiang2009} do not show any signatures of the LRO down to 2~K. In the following, we will apply a variety of numerical techniques to investigate the ground state and thermodynamic properties of CuNCN. Subsequently, we use the results of this analysis to address the sensitivity of different methods to the anticipated long-range magnetic order, and suggest a consistent interpretation for the available experimental data.

\subsection{Ground state}
\label{sec:theo}
According to Ref.~\onlinecite{tsirlin2010}, the microscopic magnetic model of CuNCN entails AFM spin chains along the $c$ direction. The chains are coupled by the nearest-neighbor ferromagnetic (FM) interaction $J_1\simeq -500$~K and the next-nearest-neighbor AFM interaction $J_2\simeq 100$~K along $a$. Additionally, the diagonal coupling $J_{ac}\simeq J_2$ connects the chains in the $ac$ plane, whereas a very weak AFM coupling $J_b\simeq 5$~K links the neighboring layers in a frustrated manner (see the bottom panels of Fig.~\ref{fig:structure}). The spin lattice is quasi-one-dimensional, yet showing a strong anisotropy of the interchain couplings along the $a$ and $b$ directions. Since $J_b$ is at least an order of magnitude smaller than the couplings along $a$, we start with analyzing the two-dimensional (2D) spin lattice in the $ac$ plane.

The couplings $J$, $J_1$, and $J_{ac}$ favor the columnar AFM order featuring parallel spins along $a$ and 
antiparallel spins along $c$ (Fig.~\ref{fig:structure}). However, the columnar order is frustrated by $J_2$ that prefers antiparallel arrangement of next-nearest-neighbor spins along $a$. To find the classical ground state of this magnetic model, we write down the energy as a function of $k_x$ and $k_z$, the components of an arbitrary propagation vector $\mathbf k=(k_x,k_z)$ in the $ac$ plane. The energy per the unit cell of the spin lattice is:
\begin{align}
  E=\frac12[ J\cos k_z& +J_1\cos k_x+J_2\cos(2k_x)+ \notag\\ 
  & +2J_{ac}\cos k_x\cos k_z)].
\end{align}
The energy minimum is at $\mathbf k=(0,\pi)$ for \mbox{$J_2/(-J_1+2J_{ac})<\frac14$} (columnar AFM phase) and at $\mathbf k=\left[\text{arccos}\left[(-J_1+2J_{ac})/4J_2\right],\pi\right]$ for \mbox{$J_2/(-J_1+2J_{ac})>\frac14$} (spiral phase). Comparing this result to the better studied model of coupled frustrated spin chains (FM $J_1$, AFM $J_2$, AFM $J$, and $J_{ac}=0$),\cite{zinke2009} we note a similar competition between columnar (collinear) and spiral phases that are separated by a quantum critical point at $J_2/J_1=-\frac14$. The diagonal coupling $J_{ac}$ shifts the quantum critical point to $J_2/(-J_1+2J_{ac})=\frac14$ without changing the nature of the competing phases. Note that the frustration is only present along the $a$ direction, hence the leading coupling $J$ along $c$ does not influence the stability of the collinear and spiral phases.

The spin lattice of CuNCN with $J_2/(-J_1+2J_{ac})\simeq 0.14$ clearly favors the columnar AFM phase. While the classical model suggests the ordered moment of 1~$\mu_B$ independent of individual exchange couplings, quantum effects have to be taken into account for a \mbox{spin-$\frac12$} system. To ensure the reliable treatment of such effects, we use the coupled-cluster method (CCM). An efficient application of CCM to low-dimensional spin systems has been illustrated in Refs.~\onlinecite{zinke2009,ccm1,ccm2,darradi2005,richter2007,darradi2008,zinke2008,bishop2008,richter2010}.
Here we give only an outline of the main relevant features of the CCM.

The core of the CCM method is the choice of a normalized reference state $|\Phi\rangle$ together with a complete set of (mutually commuting) multi-configurational creation operators $\{ C_L^+ \}$ and the corresponding set of their Hermitian adjoints $\{ C_L \}$. In our case, the reference state $|\Phi\rangle$ is the columnar AFM phase of the classical spin model (parallel spins along $a$, antiparallel spins along $c$). Then we perform a rotation of the local axis of the spins such that all spins in the reference state are directed along the negative $z$ axis, i.e., in the rotated coordinate frame the reference state reads 
$|\Phi\rangle \hspace{-3pt} = \hspace{-3pt}
|\hspace{-3pt}\downarrow\rangle |\hspace{-3pt}\downarrow\rangle
|\hspace{-3pt}\downarrow\rangle \ldots \,$.
The corresponding multispin creation operators then can be written as $C_I^+=s_\alpha^+,\,\,s_\alpha^+s_{\beta}^+,\,\,
s_\alpha^+s_{\beta}^+s_{\gamma}^+,\cdots$, where the indices $\alpha,\beta,\gamma\dots$  label arbitrary lattice sites.

The ket- and bra- ground states are given by:
\begin{eqnarray}
\label{eq_ccm}
|\Psi\rangle = e^S|\Phi\rangle, 
\qquad 
S = \sum_{I \neq 0}{\cal S}_IC_I^+ ; 
\nonumber\\
\langle\tilde{\Psi}| =  \langle\Phi|\tilde{S}e^{-S},
\qquad 
\tilde{S} = 1 + \sum_{I \neq 0}\tilde{\cal S}_IC_I .
\end{eqnarray}
Using $\langle\Phi|C_I^+=0=C_I|\Phi\rangle$ $\forall I\neq 0$, $C_0^+\equiv 1$, the commutation rules $[C_L^+,C_{K}^+] = 0=[C_L,C_{K}]$, the orthonormality condition $\langle\Phi|C_IC_J^+|\Phi\rangle=\delta_{IJ}$, and completeness $\displaystyle\sum_I C_I^+|\Phi\rangle\langle\Phi|C_I=1=|\Phi\rangle\langle\Phi|+\sum_{I\neq 0}C_I^+|\Phi\rangle\langle\Phi|C_I$, we get a set of non-linear and linear equations for 
the correlation coefficients ${\cal S}_I$ and $\tilde{\cal S}_I$, respectively. 
The order parameter (sublattice magnetization or ordered moment, $m$) is given by 
\begin{equation}
  m =-\dfrac{1}{N}\sum_{i,n}^N \langle\tilde\Psi|s_{i,n}^z|\Psi\rangle,
\end{equation}
where $s_{i,n}^z$ is the spin operator expressed in the rotated coordinate system, and $(i,n)$ denotes the position of the lattice site. The CCM provides results for infinite lattices. The only approximation of the CCM is the truncation of the expansion for the correlation operators $S$ and $\tilde S$. We use the well-established LSUB$n$ scheme that includes all multispin correlations on the lattice with $n$ or fewer neighboring sites. To account for all multispin correlations, the CCM estimates obtained on different levels of the LSUB$n$ approximations ($m_n$) are then extrapolated to the $n\rightarrow\infty$ limit ($m_{\infty}$) using the extrapolation schemes $m_n=m_{\infty}+a/n+b/n^2$ (scheme 1) and $m_n=m_{\infty}+a/n^{1/2}+b/n^{3/2}$ (scheme 2). The former scheme is typically used for quantum spin systems with the AFM long-range order,\cite{ccm1,ccm2,darradi2005,richter2007} whereas the latter scheme has been successfully applied to systems near a quantum critical point, where the magnetic order parameter $m$ is small.\cite{darradi2008,bishop2008,richter2010}

\begin{figure}
\includegraphics{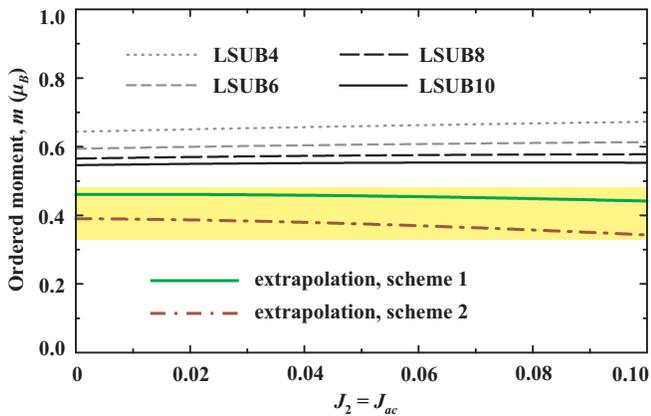}
\caption{\label{fig:ccm}
(Color online) CCM LSUB$n$ results as well as extrapolated ($n\rightarrow\infty$) data for the ordered moment $m$ at $J_1/J=-0.2$. The couplings $J_2=J_{ac}$ are given in units of $J=1$. The two extrapolation schemes are explained in the text. The shaded bar denotes the likely range for the ordered magnetic moment in CuNCN.
}
\end{figure}

\begin{figure}
\includegraphics{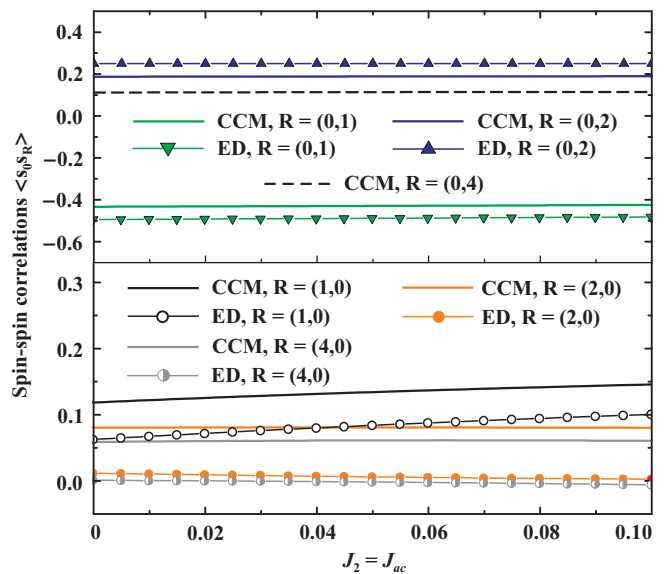}
\caption{\label{fig:sisj_a}
(Color online) CCM-LSUB8 and ED results for the spin-spin correlations $\langle {\bf s}_{0}{\bf s}_{\bf R}\rangle$ along the $c$ (top panel) and $a$ (bottom panel) directions at $J_1/J=-0.2$. The couplings $J_2=J_{ac}$ are given in units of $J=1$. The ED results for ${\bf R}=(0,4)$ are not available because of the limited size of the finite lattice.
}
\end{figure}

Fig.~\ref{fig:ccm} shows the extrapolated ordered moment $m=m_{\infty}$ for $J_1/J=-0.2$ and variable $J_2=J_{ac}$. We find that the frustrating coupling $J_2$ has no appreciable effect on the ordered moment $m$ in the relevant parameter range. Depending on the extrapolation scheme used, we find $m=0.34-0.46$~$\mu_B$, where the largest value is obtained  for scheme 1 at $J_2=J_{ac}=0$ and the lowest value is found for scheme 2 at $J_2=J_{ac}=0.5J_1$. These values are well below the classical value of 1~$\mu_B$ and also significantly lower than the ordered moment of $\sim$0.6~$\mu_B$ on the \mbox{spin-$\frac12$} square lattice (see, for example, Refs.~\onlinecite{sandvik1997,ccm2,richter2007}). This small value of the order parameter can be attributed to the dominating exchange coupling $J$ along the $c$ direction that leads to a quasi-one-dimensional system of weakly coupled antiferromagnetic spin-$\frac12$ Heisenberg chains. Simple models of such weakly coupled chains are well studied, e.g., in Refs.~\onlinecite{zinke2008,affleck94,wang97,sandvik1999,kim2000}. These works show that even an infinitesimally small (non-frustrated) interchain coupling leads to the magnetic long-range order at zero temperature. Although our model does not
correspond directly to these simple models considered in the previous literature, a similar conclusion is also valid for our model, in particular, since our numerical simulations indicate that the effect of the frustrating coupling $J_2$ is very weak.  

To get additional insight into the magnetic ordering and the role of the frustrated coupling $J_2$, we present spin-spin correlation functions $\langle {\bf s}_{0}{\bf s}_{\bf R}\rangle$ for various lattice vectors ${\bf R}$ (Fig.~\ref{fig:sisj_a}). In addition to the CCM data, we also show the results of Lanczos exact diagonalization (ED) for a $8\times 4$ finite lattice with periodic boundary conditions.\footnote{Here, we take 8 sites along $a$ and 4 sites along $c$ to account for the long-range nature of $J_2$.} Obviously, the spin-spin correlations support the anticipated columnar AFM ground state with a large negative correlation for nearest neighbors along $c$ and a sizable positive correlation for nearest neighbors along $a$. Most correlations are insensitive to $J_2$ and $J_{ac}$, although the correlation on the $J_1$ bond is slightly enhanced as $J_2$ and $J_{ac}$ increase. This signifies a slight enhancement in the coupling along the $a$ direction, yet this effect is countered by the increasing frustration. Overall, the weak changes in the spin-spin correlations are consistent with the nearly constant value of the ordered moment, as evaluated by CCM (Fig.~\ref{fig:ccm}). Note that the weak correlations along the $a$ direction support the picture of weakly-coupled spin chains leading to a small ordered moment. Although the ED and CCM results for spin-spin correlations show very similar qualitative behavior, the quantitative differences are obvious. These differences can be attributed to еру strong finite-size effects in the ED data, since in the $c$ direction we have a spin ring of 4 sites, only.

Using the CCM and exact diagonalization results, we conclude that the spin lattice of CuNCN develops the columnar AFM order in the $ac$ plane. This order corresponds to the propagation vector $\mathbf k=(0,k_y,\pi)$ with respect to the unit cell of the spin lattice (Fig.~\ref{fig:structure}, bottom right panel) or $\mathbf k=(0,k_y,0)$ with respect to the unit cell of the atomic structure. Now, we comment on the possible interlayer ordering along the $b$ direction. According to the left panels of Fig.~\ref{fig:structure}, the weak interlayer coupling $J_b$ forms isosceles triangles and therefore frustrates the interlayer order on the classical level. This classical degeneracy is lifted by quantum fluctuations favoring collinear ground states via the order-from-disorder mechanism.\cite{henley1989} We thus expect the collinear order in CuNCN, although the periodicity along $b$ cannot be determined unambiguously. Note that a similar case of weak and frustrated interlayer couplings has been reported for the Heisenberg model on the body-centered tetragonal lattice.\cite{yildirim1996} An elaborate spin-wave study of Ref.~\onlinecite{yildirim1996} suggests two collinear states with the doubled and quadrupled periodicity along the interlayer direction. Such states are strongly favored over any non-collinear configurations, yet the preference for the state with the quadrupled periodicity appears in high orders of spin-wave theory, only. Therefore, we expect the doubled or quadrupled periodicity along the $b$ direction in CuNCN, i.e., propagation vectors are $\mathbf k=0$ or $\mathbf k=(0,\pi,0)$ with respect to the crystallographic unit cell of CuNCN. 

\subsection{Thermodynamic properties}

\begin{figure}
\includegraphics{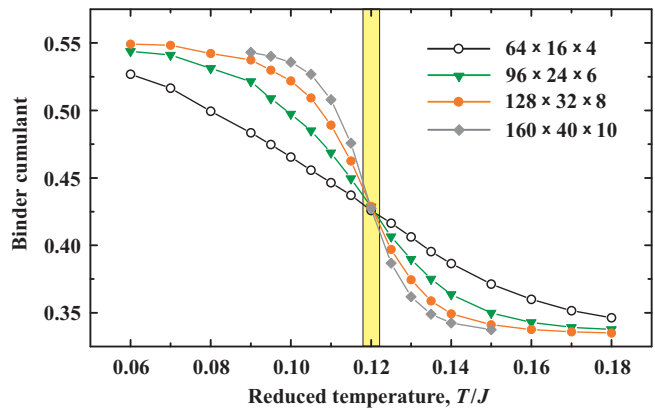}
\caption{\label{fig:Binder}
(Color online) Binder cumulant $\langle m_s^4\rangle/\langle m_s^2\rangle^2$ calculated for different system sizes at $J_1/J=-0.2$, $J_{ac}/J=0.04$, and the effective interlayer coupling $J_b^{\eff}/J=0.002$. The shaded bar shows the estimated $T_N/J\simeq 0.12$.
}
\end{figure}

CCM is a handy tool for studying frustrated spin systems, but its applications are restricted to ground-state properties. To investigate the finite-temperature behavior of our microscopic model, we have to apply ED that puts severe restrictions on the accessible lattice size (note, for example, the pronounced finite-size effects for the spin-spin correlations in Fig.~\ref{fig:sisj_a}). However, the ground-state properties are weakly dependent on $J_2$ and $J_{ac}$. Therefore, we can safely simplify our model by removing the coupling $J_2$ and eliminating the frustration. This simplification enables us to apply quantum Monte-Carlo (QMC) techniques that readily treat large systems, thus effectively overcoming finite-size effects. Since the 2D model in the $ac$ plane maintains the long-range order at zero temperature only, we also introduce a realistic interlayer coupling $J_b^{\eff}/J=0.002$ that should stabilize the LRO above 0~K. 

The spin lattice of CuNCN is spatially anisotropic, with the leading coupling along $c$, the five times weaker coupling along $a$, and a diminutively small coupling along~$b$. Following Refs.~\onlinecite{sandvik1999,sengupta2003,tsirlin2011}, we adjust the aspect ratio of finite lattices to account for different correlation lengths along $a$, $b$, and $c$. Specifically, we used $L/4\times L/16\times L$ lattices with $L\leq 160$ (up to 64000 sites). The N\'eel temperature $T_N$ is determined from the standard scaling procedure for the Binder ratio $B=\langle m_s^4\rangle/\langle m_s^2\rangle^2$, where $m_s$ is the staggered magnetization. Since $B(T)$ is independent of the lattice size at $T=T_N$,\cite{binder} the N\'eel temperature is precisely determined as the crossing point of the $B(T)$ curves calculated for different $L$. Fig.~\ref{fig:Binder} shows the scaling procedure and the resulting $T_N/J\simeq 0.12$. Using the experimental estimate of $J\simeq 2300$~K, we arrive at $T_N\simeq 275$~K, which is much larger than the experimental value of 70~K, because the non-frustrated interlayer coupling $J_b$ is considered. In CuNCN, the frustrated couplings $J_b$ will impede the LRO, thus reducing $T_N$ and bringing it closer to the experimental value.

\begin{figure}
\includegraphics{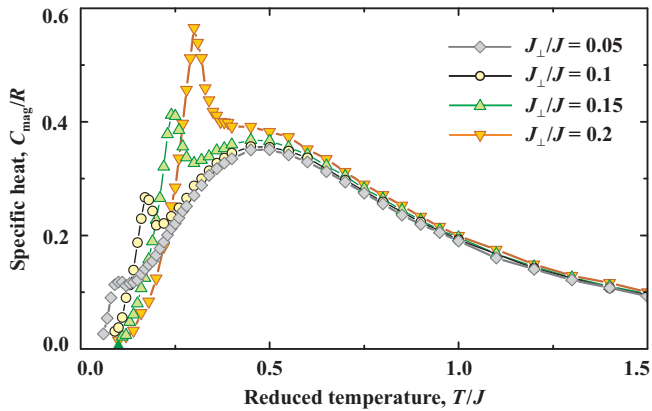}
\caption{\label{fig:heat-1}
(Color online) Magnetic specific heat of spin chains with a uniform interchain coupling $J_{\perp}/J$ of 0.05, 0.1, 0.15, and 0.2. The decrease in $J_{\perp}/J$ shifts the transition anomaly to lower temperatures and dramatically reduces its magnitude.
}
\end{figure}
\begin{figure}
\includegraphics{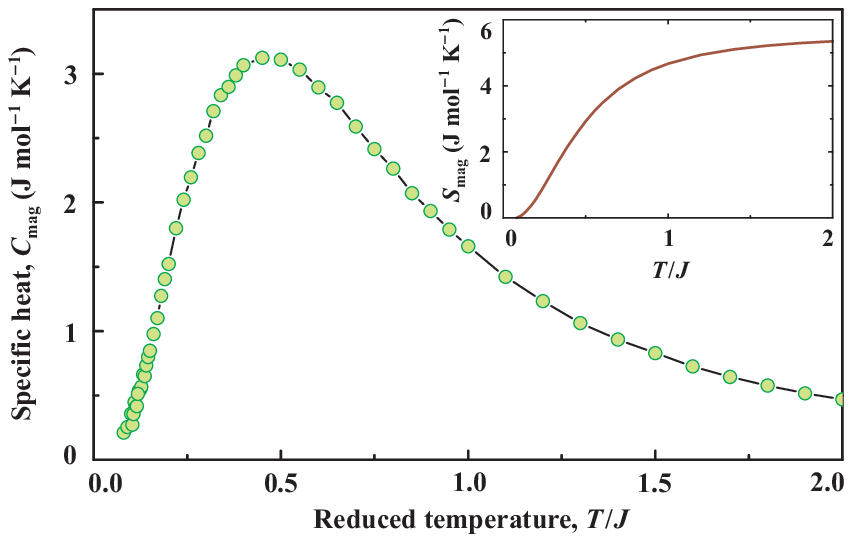}
\caption{\label{fig:heat-2}
(Color online) Magnetic specific heat for the spin lattice of CuNCN, with $J_1/J=-0.2$, $J_{ac}/J=0.04$, and $J_b^{\eff}/J=0.002$. The inset shows the magnetic entropy $S_{\mg}$ obtained by integrating $C_{\mg}/T$.
}
\end{figure}
It is instructive to compare the above numerical result to previous theoretical studies of coupled spin chains.\footnote{%
While theoretical studies address the AFM case, the leading interchain coupling in CuNCN is ferromagnetic. At this point, we do not distinguish between ferromagnetic and antiferromagnetic interchain couplings. Our simulations show that for weak interchain couplings the transition temperatures and the magnetic specific heat marginally depend on the sign of $J_{\perp}$.} Despite the abundance of spatially anisotropic interchain couplings in model quasi-1D compounds,\cite{johannes2006,janson2009,janson2010}  theoretical works rather focus on the case of isotropic interchain couplings (same coupling along $a$ and $b$),\cite{sandvik1999,yasuda2005,schulz1996} because systems with large spatial anisotropy are quite difficult to model. To facilitate the comparison, we set the diagonal coupling $J_{ac}$ to zero and consider a simplistic model of uniform chains coupled with $J_1/J=-0.2$ and $J_b^{\eff}/J=0.002$ along $a$ and $b$, respectively. The resulting $T_N/J$ is 0.09 compared to 0.12 in the model including $J_{ac}$, and shows the non-negligible role of $J_{ac}$ in enhancing the interchain coupling within the $ac$ plane. We now compare this estimate to the results of Ref.~\onlinecite{yasuda2005} for isotropic interchain couplings $J_{\perp}$. Using the averaged coupling of $J_{\perp}^{\eff}=(|J_1|+J_b)/2\simeq 0.1J$, one finds a much larger $T_N/J\simeq 0.169$.\cite{yasuda2005} This discrepancy is often bypassed by assuming an empirical three-fold or four-fold reduction in $T_N$ due to the spatial anisotropy of the interchain couplings.\cite{rosner1997,kaul2003} Our numerical results suggest that this empirical recipe underestimates $T_N/J\simeq 0.09$, which is only twice reduced compared to the estimate from the averaged interchain coupling $J_{\perp}^{\eff}$.

To explore the transition anomaly at $T_N$, we evaluated the magnetic specific heat for coupled spin chains. We first consider the isotropic interchain coupling $J_{\perp}$ (Fig.~\ref{fig:heat-1}).\footnote{%
The simulations are done for $8\times 8\times 32$ ($J_{\perp}/J=0.1,0.15$) and $12\times 12\times 24$ ($J_{\perp}/J=0.2$) finite lattices that are sufficient to avoid finite-size effects in the temperature range under investigation.} At $J_{\perp}/J=0.2$, the pronounced transition anomaly is superimposed on the broad specific heat maximum. At lower $J_{\perp}/J$, the $T_N$ is reduced, thereby shifting the anomaly below the maximum and shrinking its size. This trend follows recent experimental\cite{lancaster2007} and theoretical\cite{sengupta2003} findings for quasi-two-dimensional spin systems, where the transition anomaly is dramatically reduced and eventually becomes invisible at sufficiently weak interlayer couplings. 

In CuNCN, the very weak coupling along the $b$ direction enhances quantum fluctuations and reduces $T_N$ compared to the scenario of the averaged interchain coupling $J_{\perp}^{\eff}=(|J_1|+J_b)/2\simeq 0.1J$. This leads to a further decrease in the transition anomaly. Fig.~\ref{fig:heat-2} shows the simulated specific heat for the relevant exchange parameters of CuNCN, as derived from the DFT calculations.\cite{tsirlin2010} We were unable to discern any signatures of the transition anomaly at $T_N/J\simeq 0.12$, because the magnetic entropy available at the N\'eel temperature is exceedingly low. We illustrate this effect in the inset of Fig.~\ref{fig:heat-2}, where the magnetic entropy $S_{\mg}(T)=\int\limits_0^T\frac{C_{\mg}}{T}\,dT$ is shown. The $S_{\mg}(T_N)\simeq 0.13$~J mol$^{-1}$~K$^{-1}$ is less than 3~\% of the total magnetic entropy $R\ln 2\simeq 5.75$~J~mol$^{-1}$~K$^{-1}$. Considering the inevitable effect of exchange anisotropy (broadening of the transition anomaly) and the large lattice contribution of about 20~J~mol$^{-1}$~K$^{-1}$, which is 150 times larger than $S_{\mg}(T_N)$ (Fig.~\ref{fig:heat}), we conclude that the specific heat anomaly at the magnetic transition in CuNCN can not be observed experimentally, especially in experiments on polycrystalline samples.

\begin{figure*}
\includegraphics{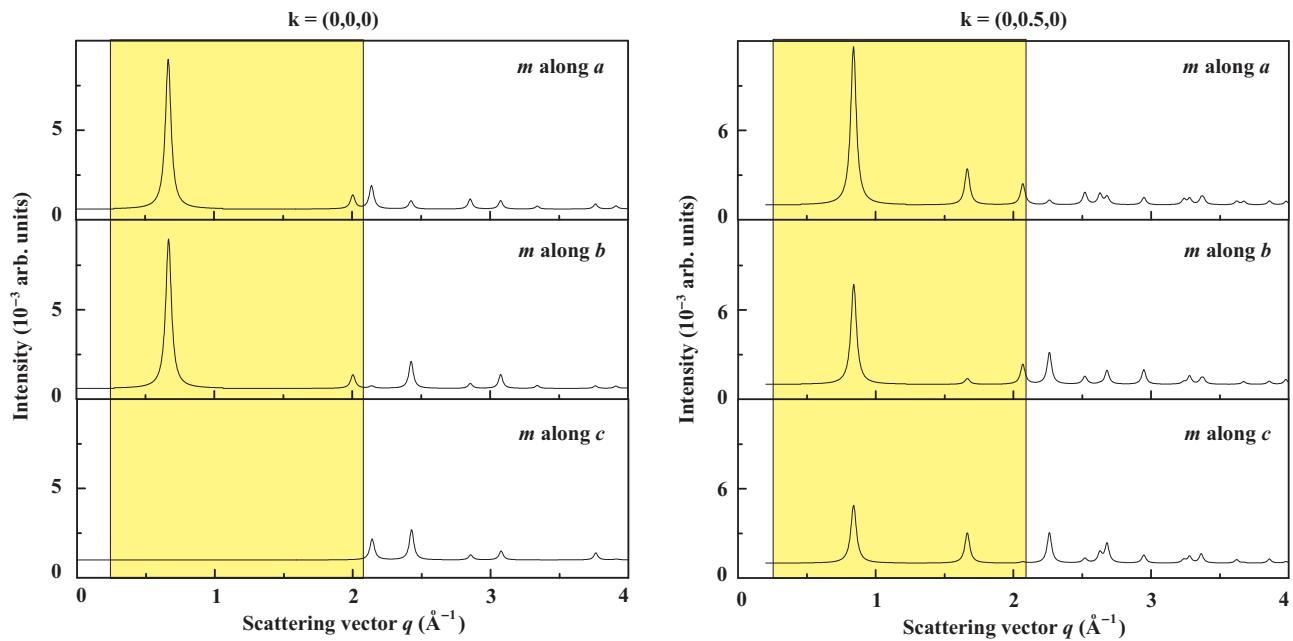}
\caption{\label{fig:neutrons}
(Color online) Simulated magnetic scattering from CuNCN for the possible propagation vectors $\mathbf k=0$ (left panel) and $\mathbf k=(0,\frac12,0)$ (right panel) and different directions of the magnetic moment. The shaded area denotes the $q$ range probed in the polarized-scattering experiment of Ref.~\onlinecite{xiang2009}. The intensity of the largest nuclear reflection is scaled to 1 arb. unit. Note that the magnetic scattering is at least 100 times weaker than the nuclear scattering.
}
\end{figure*}
Altogether, our QMC simulations are restricted to the simplified, non-frustrated spin lattice and may not be sufficient for the quantitative description of CuNCN. Nevertheless, we are able to demonstrate that the expected N\'eel temperature is quite low, and the magnetic specific heat does not show any signatures of the magnetic transition. In CuNCN, the frustrated arrangement of the interlayer couplings $J_b$ will further impede the magnetic order, thus reducing $T_N$ and leaving no room for the experimental observation of the magnetic transition in heat-capacity data.

\subsection{Neutron scattering}

The intensity of the neutron scattering from a magnetic structure is proportional to the square of the magnetic structure factor, which is, in turn, proportional to the ordered magnetic moment. Therefore, for low magnetic moments the intensity of magnetic reflections in a neutron diffraction experiment decreases dramatically. Previous studies\cite{liu2008,xiang2009} put forward the lack of the magnetic scattering as one of the arguments against the LRO in CuNCN. To find out whether or not the anticipated LRO could be observed in these experiments,\cite{liu2008,xiang2009} we simulated neutron diffraction patterns of CuNCN. The input parameters are the propagation vector and the ordering pattern, the ordered moment and its direction, as well as the scale factor for the atomic structure. The latter is necessary to scale the magnetic reflections with respect to the nuclear peaks. The ordering pattern in the $ac$ plane and the ordered moment are provided by our theoretical study in Sec.~\ref{sec:theo}. Since the direction of the magnetic moment is presently unknown, different possibilities were explored. We used the propagation vectors $\mathbf k=0$ and $\mathbf k=(0,\pi,0)$, as explained in Sec.~\ref{sec:theo}.

Simulated patterns for two possible propagation vectors and three representative directions of the magnetic moment are shown in Fig.~\ref{fig:neutrons}. The maximum intensity of the magnetic scattering is about 1.2~\% (hereinafter, we measure intensities as fractions of the largest nuclear peak) for the magnetic reflection at $q=0.6-0.8$~\r A$^{-1}$ ($001+\mathbf k$). At higher $q$ values, the magnetic intensities are notably lower because of the rapidly decreasing magnetic form factor of Cu$^{+2}$. However, even the strong magnetic reflection at low $q$ may be hard to observe. For example, Fig.~1 of Ref.~\onlinecite{liu2008} presents the low-temperature neutron-diffraction data for CuNCN. In these data, the low-angle 002 nuclear reflection with a sizable intensity of 2.3~\% is fully masked by the background. This implies that the data are by far insufficient to observe the magnetic scattering in CuNCN for any of the ordering patterns.

The polarized-scattering data of Ref.~\onlinecite{xiang2009} have an additional advantage of detecting the magnetic reflections in a separate channel, thereby probing the possible magnetic contributions to nuclear peaks and reducing the background. An inevitable drawback is the reduction in the total neutron flux because of the polarization filter. Therefore, the observation of the magnetic scattering from a spin-$\frac12$ magnet with the low magnetic moment of about 0.4~$\mu_B$ is at best challenging, at least with the instrument used in Ref.~\onlinecite{xiang2009}.\footnote{%
The DNS instrument at FRM II, as used in Ref.~\onlinecite{xiang2009}, has a flux of $5\times 10^6-10^7$~n~cm$^{-2}$~s$^{-1}$ in the polarized mode. This is comparable to the typical flux at high-resolution diffractometers, e.g., D2B at ILL when operating in the high-intensity mode. However, specially designed instruments, such as D20 at ILL, can reach an order-of-magnitude higher flux of nearly $10^8$~n~cm$^{-2}$~s$^{-1}$, which is essential to detect magnetic scattering from quantum spin systems with strongly reduced magnetic moments.
} We also note that the resolution in the nuclear channel was high enough to detect an imperfection in the sample, as evidenced by a foreign peak at $q=0.57$~\r A$^{-1}$ that brings back the problem of structural and chemical homogeneity of large powder samples used in neutron experiments. 

In the polarized neutron-scattering experiment, no magnetic peaks were observed up to $q_{\max}\simeq 2.1$~\r A$^{-1}$.\footnote{%
The data in Ref.~\onlinecite{xiang2009} are collected up to $q=2.25$~\r A$^{-1}$. A strong magnetic scattering observed at $q=2.15$~\r A$^{-1}$ is ascribed to a ``systematic error in the separation process'', thereby restricting the actual angle range to $q\leq 2.1$~\r A$^{-1}$.}
Even assuming that the experiment of Ref.~\onlinecite{xiang2009} were sufficiently sensitive to detect any magnetic scattering below $q_{\max}$, we find at least one ordering pattern ($\mathbf k=0$, $m$ along $c$) yielding magnetic reflections above $q_{\max}$ only. In this pattern, the lack of the intense reflection at $q\simeq 0.65$~\r A (001 nuclear peak) is because of symmetry restrictions, since the magnetic moment along $c$ forbids the magnetic contribution to $00l$. Therefore, the polarized-scattering experiment reported in Ref.~\onlinecite{xiang2009} can not be considered as an ultimate test for the presence or absence of LRO in CuNCN. Neutron diffraction measurements for CuNCN largely remain an open problem, and require more sensitive instruments accessing larger angle range or, preferably, the use of single crystals.

\section{Discussion}
\label{sec:discussion}

We have shown that CuNCN is a low-dimensional magnet featuring strong quantum fluctuations and an intricate magnetic transition around 70~K. In the following, we propose a plausible microscopic scenario for the magnetism of this compound, and start with analyzing the magnetic state of Cu$^{+2}$. Previous computational studies controversially reported a magnetic\cite{tsirlin2010} as well as a non-magnetic\cite{xiang2009} state for copper. This issue is now resolved by the ESR experiment showing a characteristic absorption line that signals the presence of the localized magnetic moment (Fig.~\ref{fig:esr}). The localized behavior is further underpinned by the increase in the magnetic susceptibility above room temperature (Fig.~\ref{fig:fit}). Our fit to the data suggests a very strong exchange coupling of about 2300~K. These observations are in remarkable agreement with the computational predictions of Ref.~\onlinecite{tsirlin2010}, and identify CuNCN as a charge-transfer insulator that complements the broad family of Cu$^{+2}$-based insulating quantum magnets.

A puzzling feature of CuNCN is the magnetic transition around 70~K. This transition was first evidenced by the bend in the temperature dependence of the magnetic susceptibility, but the lack of the corresponding specific heat anomaly as well as the missing magnetic scattering in the neutron diffraction data led the authors of Refs.~\onlinecite{liu2008,xiang2009} to conclude on the extrinsic nature of the susceptibility feature. In Sec.~\ref{sec:ground}, we have shown that neither specific heat nor neutron diffraction measurements are sufficiently sensitive to the possible LRO in CuNCN. Further, our $\mu$SR data reveal the formation of static magnetic field in the bulk of the stoichiometric CuNCN sample below 70~K. The emergence of static fields is accompanied by a vanishing absorption line in the ESR spectrum, thus confirming the onset of the LRO.

Experimental techniques have different sensitivity to the LRO magnetic state and may additionally involve perturbations, such as interactions with implanted muons. Therefore, it is important to examine all possible scenarios that could explain the experimental data. The vanishing ESR absorption is typical for both LRO and spin-gap states of a material. While the gapped singlet state does not feature any static fields, it is fragile toward interactions with muons that show sizable depolarization characteristic of a local spin-glass state in the vicinity of the implanted muon. Such states are characterized by the zero internal field and the monotonous decrease in the muon polarization asymmetry $A(t)$.\cite{andreica2000,*fudamoto2002,*aczel2007} Our data show the dip in $A(t)$ that signifies the non-zero internal field (Fig.~\ref{fig:muSR_FT}) and oscillations typical for the LRO state. Moreover, the dynamic relaxation rate $\Lambda$ shows the anomaly at $T_N$, which is also consistent with the LRO scenario. The observation of the non-zero internal field excludes the formation of the muon-driven spin glass in CuNCN. The spin gap in CuNCN would further contradict the microscopic analysis (Sec.~\ref{sec:ground} and Ref.~\onlinecite{tsirlin2010}), because the proposed spin lattice leaves no room for the spin gap formation.

An opposite situation is known for gapless quantum magnets, where muons disrupt exchange pathways and induce a local magnetic order that results in well-defined oscillations of $A(t)$ and few sharp peaks in the Fourier-transformed spectrum.\cite{chakhalian2003,*storchak2009} This is again different from our observations (Figs.~\ref{fig:muSR_t_spectra} and~\ref{fig:muSR_FT}). More importantly, the muon-induced magnetic order has its own temperature scale that is usually different from the intrinsic temperature scale of the system.\cite{chakhalian2003,*storchak2009} In CuNCN, the onset of static magnetic fields, as probed by muons, is accompanied by the vanishing ESR absorption. Therefore, the effect is clearly intrinsic. At this point, we mention the recent results by Zorko \textit{et al.},\cite{zorko} who also observed the strong muon depolarization and the vanishing ESR line below 70~K. Zorko \textit{et al.}\cite{zorko} ambiguously attribute the $\mu$SR signal to a spin-glass state induced by the implanted muon, although their spectra show a similar dip in $A(t)$ and likely lead to the non-zero internal field. Unfortunately, the authors of Ref.~\onlinecite{zorko} do not consider the possibility of the LRO state, which is the most likely and microscopically justified scenario for the ground state of CuNCN.

We have shown that the available experimental data on CuNCN are consistently explained in the framework of the microscopic scenario proposed in Ref.~\onlinecite{tsirlin2010}. The quasi-1D spin lattice leads to strong quantum fluctuations that reduce the N\'eel temperature and the ordered moment, thus impeding the experimental observation of the columnar AFM LRO by heat-capacity measurements and neutron scattering. The results of the magnetic susceptibility, $\mu$SR, and ESR are a compelling evidence for the emergence of static magnetic fields below 70~K, in agreement with the developed microscopic scenario. The onset of static magnetic fields is further corroborated by the broadening of the $^{14}$N nuclear magnetic resonance (NMR) line reported by Zorko \textit{et al.}\cite{zorko} 

Despite this robust qualitative scenario, several aspects of the low-temperature magnetic behavior require further attention. One obvious problem is the broad distribution of internal fields probed by the implanted muons (Fig.~\ref{fig:muSR_FT}). This broad distribution is rather unusual for the LRO state that typically leads to one or few characteristic fields at muon stop sites, well-defined oscillations, and narrow peaks in the Fourier-transformed spectra. A plausible explanation is a large number of inequivalent stop sites with different internal fields. For example, the $\mu$SR study of spin-$\frac12$ frustrated square lattices in V$^{+4}$ phosphates\cite{carretta2009} revealed a similarly broad distribution of internal fields in Pb$_2$VO(PO$_4)_2$ (the variance of the distribution is about 10~mT, as compared to $20-25$~mT in CuNCN), although the LRO in this compound is independently confirmed by polarized neutron scattering.\cite{skoulatos2009} The crystal structure of CuNCN is quite simple, but its relation to the possible muon stop sites is far from obvious, because the negative charge is associated with rather delocalized molecular orbitals of the NCN unit,\cite{tsirlin2010} in contrast to the localized oxygen orbitals in oxides. A further microscopic analysis of the possible muon stop sites is highly desirable to achieve more complete understanding of the $\mu$SR response in CuNCN and other transition-metal carbodiimides. 

A closely related problem is the broad distribution of NMR spin-lattice relaxation times reported in Ref.~\onlinecite{zorko}. This finding is based on fitting the magnetization recovery with a stretched exponent, and taken as an evidence for an inhomogeneous magnetic ground state of CuNCN. Fitting the magnetization recovery with a stretched exponent produces a cumulative parameter $\beta$ that should be equal to 1 for the single relaxation time. The deviation from unity either describes a broad distribution of relaxation times or the presence of several distinct relaxation times in the system under investigation.\cite{johnston2006} For example, the strongly reduced $\beta\simeq 0.6$ (compare to $\beta\simeq 0.5$ in CuNCN\cite{zorko}) has been observed in the LRO state of CaV$_2$O$_4$.\cite{zong2008} However, the magnetic ordering in CaV$_2$O$_4$ is in no way inhomogeneous, as shown by the comprehensive neutron-scattering study.\cite{niazi2009} Therefore, the complex behavior of the magnetization recovery in CuNCN could also be understood as the presence of several distinct relaxation times in the LRO state.

The last, and probably most severe problem is the smeared magnetic transition in CuNCN. We identify $T_N\simeq 70$~K as the onset temperature, although the complete magnetic order in the bulk of the CuNCN sample is established below 20~K, only (Fig.~\ref{fig:muSR_ZF_results}, see also Fig.~2 in Ref.~\onlinecite{zorko}). The broad intermediate region between 20 and 70~K is somewhat unusual for crystalline magnetic systems showing abrupt (and typically second-order) transitions to the LRO state. The broadening of the transition might be related to the peculiar microstructural effects observed in our synchrotron XRD experiment (Fig.~\ref{fig:xrd}). 

Altogether, we argue that the bulk of the experimental observations on CuNCN can be reconciled within the quasi-1D microscopic scenario proposed in Ref.~\onlinecite{tsirlin2010}. The Cu$^{+2}$ sites hold localized magnetic moments, and the temperature of 70~K is the characteristic temperature related to the onset of the LRO state with the strongly reduced magnetic moment of about 0.4~$\mu_B$. However, some of the previously reported experimental results require further consideration. For example, more sensitive neutron-scattering experiments would be essential to detect or refute the LRO in CuNCN. Although the $\mu$SR and NMR response might be characteristic of a partial inhomogeneity, the origin of this effect could be purely magnetic (e.g., the formation of a spin-glass state), purely structural, or both. We have shown that even stoichiometric samples of CuNCN feature a sizable amount of defects that may affect the $\mu$SR and NMR signals. Another interesting result is the variable chemical composition of CuNCN and the accommodation of foreign elements, such as oxygen and hydrogen, within the same structure type. Our study of the non-stoichiometric sample 2 reveals the enhanced inhomogeneity of the magnetic state probed with $\mu$SR and at least a partial destruction of the LRO state. Therefore, intentional deviations from the ideal stoichiometry may be a feasible way to approach the fully inhomogeneous magnetic ground state of CuNCN, similar to the one claimed in Ref.~\onlinecite{zorko} for the stoichiometric compound.

\acknowledgments

This work was partly performed at the Swiss Muon Source (S$\mu$S), Paul Scherrer Institute (PSI, Switzerland). We are grateful to ESRF for granting the beam time for the high-resolution XRD experiments. Experimental assistance of Rita Gellrich (sample preparation), Yurii Prots and Horst Borrmann (in-house XRD), Gudrun Aufferrmann (chemical analysis), Stefan Hoffmann (thermal analysis), Caroline Curfs and Andrew Fitch (ESRF) is kindly acknowledged. We are also grateful to Artem Abakumov, Edward Levin, Matteo Leoni, Ramesh Nath, and Ulrich Schwarz for fruitful discussions and sharing our interest in CuNCN. We are indebted to Richard Dronskowski for stimulating discussions and fascinating ideas regarding CuNCN. A.T. was funded by Alexander von Humboldt Foundation. A. M. acknowledges support by the Swiss National Science Foundation and the NCCR Materials with Novel Electronic
Properties (MaNEP).

%

\end{document}